\documentclass[modern,dvipdfmx]{aastex62}
\usepackage[dvipdfmx]{graphicx}
\newcommand{\etal }{{et al.} }
\newcommand{\msun}{\thinspace M_\odot}

\newcommand{\vect}[1]{\mbox{\boldmath$#1$}}
\def\lesssim{\mathrel{\hbox{\rlap{\hbox{\lower4pt\hbox{$\sim$}}}\hbox{$<$}}}}
\def\gtrsim{\mathrel{\hbox{\rlap{\hbox{\lower4pt\hbox{$\sim$}}}\hbox{$>$}}}}
\newcommand{\cm}{\,{\rm cm}^{-3} } 
\newcommand{\km}{\,{\rm km\, s}^{-1}}

\submitjournal{ApJ}
\shorttitle{The First Two Thousand Years}
\shortauthors{M. N. Machida and S. Basu}

\begin{document}
\title{The First Two Thousand Years of Star Formation}

\correspondingauthor{Masahiro N. Machida}
\email{machida.masahiro.018@m.kyushu-u.ac.jp}

\author[0000-0002-0963-0872]{Masahiro N. Machida}
\affil{Department of Earth and Planetary Sciences, Faculty of Sciences, Kyushu University, Fukuoka 812-8581, Japan}

\author[0000-0003-0855-350X]{Shantanu Basu}
\affiliation{Department of Physics and Astronomy, The University of Western Ontario, London, ON N6A 3K7, Canada}

\begin{abstract}
Starting from a prestellar core with a size of $1.2\times10^4$\,AU, we calculate the evolution of a gravitationally collapsing core until $\sim2000$\,yr after protostar formation using a three-dimensional resistive magnetohydrodynamic simulation, in which the protostar is resolved with a spatial resolution of $5.6\times10^{-3}$\,AU. 
Following protostar formation, a rotationally supported disk is formed. 
Although the disk size is as small as $\sim2-4$\,AU, it remains present until the end of the simulation. 
Since the magnetic field dissipates and the angular momentum is then not effectively transferred by magnetic effects, the disk surface density gradually increases and spiral arms develop due to gravitational instability.
The disk angular momentum is then transferred mainly by gravitational torques, which induce an episodic mass accretion onto the central protostar. 
The episodic accretion causes a highly time-variable mass ejection (the high-velocity jet) near the disk inner edge, where the magnetic field is well coupled with the neutral gas. 
As the mass of the central protostar increases, the jet velocity gradually increases and exceeds $\sim100\km$.
The jet opening angle widens with time at its base, while the jet keeps a very good collimation on the large scale. 
In addition, a low-velocity outflow is driven from the disk outer edge. 
A cavity-like structure, a bow shock and several knots, all of which are usually observed in star-forming regions, are produced in the outflowing region.
\end{abstract}


\section{Introduction}
\label{sec:intro}
Stars are a very important constituent of the universe, and their formation proceeds through a very large dynamic range of length  and time scales. It is important to resolve the maximum possible range of scales in this process, and follow the interacting effects of the magnetic field, angular momentum, and magnetic field dissipation.
Star formation has been investigated extensively in both observational and theoretical works. 
Recently, observations have showed a dramatic progress in understanding the star formation process.  
The very early phase of star formation is being unveiled by large-sized telescopes such as ALMA. Circumstellar structures such as the infalling envelope, rotationally supported disk, and outflow driving region around very young protostars are being spatially resolved \citep[e.g.][]{sakai14,ohashi14,aso15,lefloch15, plunkett15, ching16,tokuda16,perez16,hirota17,bjerkeli16,aso17,alves17,lee17}.
The new findings by ALMA can constrain the scenario for star formation that had been constructed mainly based on theoretical works, because past observations with poor spatial resolution could not resolve the star-forming sites such as the disk-forming and outflow-driving regions.

Prior to the ALMA era, some possible scenarios for the early phase of the star formation have been offered and discussed. 
Since the very early phase of stars or very small scales around protostars could not be observed, we could not verify which scenario is most realistic.
For example, some researchers claimed that magnetic field strongly suppresses disk formation and that a Keplerian disk appears only in the very late stage of the star formation \citep{mellon08,li11,li14}, while some researchers pointed out that the dissipation of magnetic field promotes disk formation in the very early stage of star formation \citep{dapp10,inutsuka12,tsukamoto15,tsukamoto15b,wurster18,wurster18b}. 
At present, Keplerian disks are being observed around Class 0 objects \citep{tobin12,murillo13,sakai14,lee14,ohashi14,segura-Cox16,yen17}, and thus the disk formation scenario can be constrained.  
We can now directly compare observations with theoretical models or numerical simulations, and thereby construct an accurate star formation scenario.

 
Here, we simply summarize recent magnetohydrodynamic (MHD) star-formation simulations that follow the formation of a disk and outflow.
Using two-dimensional ideal MHD simulations, \citet{tomisaka98,tomisaka00,tomisaka02} investigated the protostar formation and outflow driving in gravitationally collapsing clouds, showing that outflows can be generated self-consistently from the magnetized collapse.
Subsequently, three dimensional ideal \citep{banerjee06} and non-ideal \citep{machida07} MHD simulations were used to investigate the driving mechanism of the low-velocity outflow and high-velocity jet during the gas collapsing phase, in which the cloud evolution was calculated until formation of a protostar. 
In the latter study \citep{machida07}, only Ohmic dissipation was considered as non-ideal MHD effects. 
There are three non-ideal MHD effects  (Ohmic dissipation,  ambipolar diffusion and Hall effect).
The former two (Ohmic dissipation and ambipolar diffusion) weaken magnetic fields and thus  magnetic effects, while the third one (Hall effect) changes the configuration of magnetic field lines \citep[for details, see][]{tsukamoto15,tsukamoto15b,wurster15,wurster16}.
Recently, the formation of a rotationally-supported disk in the very early accretion phase was investigated using radiation magnetohydrodynamic (RMHD) simulations that also included the non-ideal MHD effects  \citep{tomida15,tsukamoto15,vaytet18,wurster18}.
These multi-dimensional works resolved the protostar and protostellar environment with sufficient spatial resolution that one can try to directly compare them with recent high-resolution observations.

However, for these high spatial resolution simulations, there is a problem with long-term time integration after protostar formation. 
The typical timescale of self-gravitating gas is the free-fall time, which is proportional to the inverse square root of the density i.e., $\rho^{-1/2}$. 
The protostar has a density of  $\rho\gtrsim 10^{-2}$\,g\,$\cm$ \citep{larson69,masunaga00}, which corresponds to a free-fall time scale of $\lesssim 10^{-3}$\,yr. 
On the other hand, the mass accretion (i.e., star formation) lasts for $\sim10^{5-6}$\,yr, which roughly corresponds to the free-fall time scale of the initial star forming cloud. 
Thus, at least $10^{8-9}$ time steps of integration are necessary. 
Note that, in reality, the calculation time step becomes $\ll 10^{-3}$\,yr because we need to resolve the Alfv\'en velocity crossing time across cells in MHD simulations. 
Therefore, a considerably high CPU cost is required to investigate the gas accretion phase following protostar formation, when spatially resolving the protostar. 
 
Instead of resolving the protostar, a sink cell is commonly used in order to extend the time scale of integration. In this case, the gas inside the sink radius $r_{\rm sink}$ is removed and added to the gravitational potential of protostar \citep{bate95,krumholz04}. 
The sink radius is usually set to $r_{\rm sink}\gtrsim 0.1-10$\,AU. 
The sink technique enables us to investigate the long-term evolution of star-forming clouds and determine the final stellar mass or star formation efficiency \citep[e.g.,][]{machida13} and the  morphological evolution of circumstellar disk \citep[][]{tomida17}. 
However, the structure inside the sink radius is not resolved in such simulations.
On the other hand, the ALMA observations are already resolving a spatial scale of $\sim 1-10$\,AU in nearby star-forming regions \citep[e.g.,][]{alma15}. 
Thus, in order to directly compare high-resolution observations with simulations, we need to calculate the cloud evolution without a sink cell. 
Moreover, in the low-mass star formation process, a simulation with a sink cell cannot resolve the formation of a nascent disk and the high-velocity jet that is driven from near the protostar \citep{machida13,machida14b,machida16}.
Since this study focuses on the early phase of star formation or the evolution of the nascent disk and the jet driving, we do not describe in detail the results of various studies that employ a sink cell.

Although researchers could spatially resolve a nascent disk and the jet driving region in simulations without a sink, there is a difference in time between the simulations and observations.  
It is very difficult to observe the protostellar system at a very early stage, because the time spent in it is only a small fraction of the total lifetime.
Only a few protostars that have age $t\lesssim100-1000$\,yr are confirmed \citep[e.g.,][]{takahashi12a,takahashi12b}, in which the protostellar age (or time scale) was estimated from the protostellar outflow (the outflow length is divided by the outflow speed). 
On the other hand, it is difficult for a simulation to proceed beyond 1-10\,yr after protostar formation (while resolving the protostar) due to the time step limitations.
As a result, in many such simulations, the calculation was stopped within $\lesssim 1-10$\,yr after protostar formation \citep{tomisaka02,banerjee06,machida07,bate14,tomida15,tsukamoto15,vaytet18,wurster18}.

In a previous paper, using three-dimensional resistive MHD simulations, the cloud evolution was calculated for 270\,yr after protostar formation \citep[][hereafter Paper I]{machida14}. Here in this paper we study the cloud evolution for $\sim2000$\,yr after protostar formation. This allows us to reach an age for which the simulation can be directly compared with recent observations of very young systems.

Our paper is structured as follows. 
The initial conditions and numerical method are described in \S\ref{sec:method}. 
The calculation results are shown in \S\ref{sec:results}.
We discuss the disk formation and parameter dependence in \S\ref{sec:discussion}. 
A summary is presented in \S\ref{sec:summary}.

\section{Initial Condition and Numerical Method}
\label{sec:method}
The initial settings and numerical methods are the same as those described in Paper I.
The difference between Paper I and this study is in the duration of time integration.
We calculated the cloud evolution for $\sim 270$\,yr after protostar formation in Paper I, while we integrate to $\sim2000$\,yr in this study, revealing qualitatively new features of early protostellar disk and outflow evolution.
This additional cloud evolution required $\gtrsim 2$\,years of wall-clock-time  (or $\gtrsim 70,000$ hours of CPU time).
Since the initial settings and methods were described in Paper I and our previous papers, we briefly describe them in this section.

The initial state is a spherical cloud having a Bonnor-Ebert density profile, in which the central number density $n_{0}$ and isothermal temperature $T_{\rm iso}$ are adopted to be $n_{0}=6 \times 10^5 \cm$ and $T_0=10$\,K. 
The cloud has a mass $M_{\rm cl}=1.7 \msun$ and a radius $R_{\rm cl}=1.2\times10^4$\,AU. 
Note that we used twice the critical Bonnor-Ebert radius, and enhanced the cloud density by a factor of 1.5 to realize a gravitationally unstable state \citep{machida13,matsushita17}. 
Furthermore, we imposed a rigid rotation ($\Omega_0 = 1.1\times10^{-13}$\,s$^{-1}$) and uniform magnetic field ($B_0 =4.5 \times 10^{-5}$\,G) to the prestellar cloud. 
The ratio of magnitudes of rotational to gravitational energy is $\beta_0=0.02$.
The mass-to-flux ratio normalized to the critical value and estimated for the whole cloud (i.e., the region $r\leq R_{\rm cl}$) is $\mu_0=1.2$
\footnote{The initial condition is the same as in \citet{machida14}, but the cloud mass and mass-to-flux ratio are written incorrectly in \citet{machida14}. 
The values described in this paper are correct. 
}.
Since the prestellar cloud is in a magnetically supercritical state, it can collapse to form a protostar within several free-fall time scales. 

The calculation is executed with the nested grid code whose details are presented in our previous papers \citep[e.g.,][]{machida04,machida05a,machida05b,machida07,machida11c,machida18}. 
In the simulation, we solve the three-dimensional resistive MHD equations (see eqs. [1]--[4] of \citealt{machida12}) with a barotropic equation of state $P \propto \rho^{\Gamma}$, in which only Ohmic resistivity is considered as the non-ideal MHD effect.
The coefficient of Ohmic resistivity was estimated based on \citet{nakano02} and  described in \citet[][see eq.~(10)]{machida07}. 
The polytropic index $\Gamma$ is adopted to mimic  the thermal evolution acquired by radiation (magneto)hydrodynamics simulations \citep{masunaga00,tomida13}, which is  described by equation (5) of \citet{machida14}. 
Note that we adjusted the polytropic index $\Gamma$ only in high-density regions in order to reproduce the size-mass relation of the protostar \citep[for details, see][]{machida14,machida15}.

Our grid has cells ($i, j, k$) = (64, 64, 32) and mirror symmetry is used at the $z=0$ plane. 
Some levels of the grid are nested and the grid level is represented by $l$. 
A finer grid is embedded in a coarser grid, in which the cell width in the coarser grid ($l=l_c$) is twice as large as that in the finer grid ($l=l_c-1$), where $l_c$ is the level of an arbitrary grid.  
Five grid levels ($l=5$) are prepared before the beginning of the calculation.
Then, finer grids are successively generated to ensure the Truelove condition \citep{truelove97}, in which the Jeans length is resolved by at least 16 cells. 
The coarsest grid has a grid size of $L_1=3.8 \times 10^5$\,AU and a cell width of $h_1=591$\,AU, while the finest grid has $L_{21}=0.35$\,AU and $h_{21}=5.6\times10^{-3}$\,AU  ( = 1.2 R$_\odot$). 
At the end of the simulation, the maximum density reaches $3\times10^{20}\cm$. 
Thus, we cover a large difference in spatial scale and density contrast.

We impose a computational boundary at a distance 16 times larger than the cloud radius ($R_{\rm bd}=16R_{\rm cl}$), in order to suppress artificial effects such as reflection of waves at the boundary. 
This means that the prestellar cloud is embedded in a large amount of interstellar medium, which has a density of $1.1\times10^{4}\cm$.  
In addition, we impose the gravity (self-gravity plus gravity of the protostar after it forms) only inside the initial cloud radius $r<R_{\rm cl}$. 
Thus, the interstellar medium only exists to suppress boundary effects (for details, see \citealt{machida13,machida14,machida18}).

\section{Results}
\label{sec:results}
We calculated the cloud evolution from the prestellar cloud core stage until $\sim2000$\, yr after protostar formation.
The gravitational collapse occurs in the prestellar core and the first core forms.
Then, the second collapse begins in the first core and a protostar forms \citep{larson69,masunaga00}. 
We identify the cloud evolution before protostar formation as `the gas collapsing phase', and the evolution after protostar formation as `the gas accretion phase.'
Although we calculated the gas collapsing phase, we do not comment on it in this paper, and instead focus on the gas accretion phase following protostar formation. The gas collapsing was thoroughly investigated in our previous works \citep[e.g.][]{machida05a,machida05b,machida06,machida07,machida11c,machida18}.
In this section, we simply summarize the gas collapsing phase by referring to past works, and then describe the gas accretion phase with our simulation results.

\subsection{Gas Collapsing Phase and Protostar Formation}
\label{sec:disk}
In this subsection,  we overview the cloud evolution during the gas collapsing phase.  
The first hydrostatic core forms in the collapsing cloud when the central number density of the collapsing cloud exceeds $\sim10^{10}\cm$ \citep{larson69,masunaga00,tomida13}.
Inside the first core, the magnetic field effectively dissipates because the ionization degree is considerably low \citep{nakano02,machida07}.  
Thus, the magnetic braking becomes ineffective and the first core is supported partly by rotation \citep{tomida15,tsukamoto15,masson16,wurster18}. 
However, the first core is mainly supported by the pressure gradient force against gravity. 
After the number density exceeds $\sim10^{16}\cm$, the dissociation of molecular hydrogen induces the second collapse.

When the number density reaches $\gtrsim 10^{20}\cm$, a protostar forms in the center of the collapsing cloud \citep{tomida13}.
However, a large part of the first core (the first core remnant) remains around the protostar.
Since the first core remnant has an angular momentum, it evolves into the rotationally supported disk \citep{bate98,walch09,walch12}.
Note that the first core remnant does not always evolve into a rotationally supported disk.
For example, if the initial cloud has a strong magnetic field, no rotationally supported disk appears just after protostar formation \citep{machida18}.
\citet{machida11c} showed that the rotationally supported disk is formed from the first core (remnant)  with the size of $0.3-10$\,AU, using a simulation in which the evolution of  a cloud with supercritical mass-to-flux ratio was calculated for $\sim3$\,yr after protostar formation \citep[see also][]{bate11}.

\subsection{Disk Evolution}
\label{sec:disk}
Figure~\ref{fig:1} shows the time evolution of the circumstellar disk in our simulation for about $2000$\,yr after protostar formation.
\footnote{ 
It should be noted that, in two-dimensional figures, we used physical quantities (e.g., gas density and velocity) located closest to the $z=0$ (or $x=0$, $y=0$) plane, because there are no physical quantities exactly on each plane (or on each axis) in our numerical code \citep[see][]{fukuda99,matsu03a}. 
Thus, the plotted quantities are slightly detached from the $z=0$ (or $x=0$, $y=0$) plane by $h(l)/2$, where $h(l)$ is the cell with of $l$-th grid. 
A discontinuity seen in the grid boundary in two-dimensions is due to the difference in the cell width of each grid.  
}
The disk gradually increases its size during $t_{\rm ps} \lesssim 500$\,yr while the disk keeps its size of $\sim 4$\,AU for $t_{\rm ps} \gtrsim 500$\,yr. Here $t_{\rm ps}$ is the elapsed time after protostar formation, in which we defined the protostar formation epoch $t_{\rm ps}=0$ when the central number density reaches $n_{\rm c}=10^{18}\cm$.
The disk size roughly corresponds to the size of the magnetically inactive region, where the magnetic field has dissipated and is not coupled with the neutral gas.
Although gas from the infalling envelope with large angular momentum later falls onto the central region, the angular momentum is effectively transferred by the magnetic braking in the outer disk region where the magnetic field is well coupled. 
This magnetic braking is effective during the gas accretion phase that lasts for $\sim 10^5$\,yr, during which time the disk cannot grow. 
As shown in Figure~\ref{fig:1}, the magnetic braking suppresses the disk growth and limits the disk size within $\lesssim 5$\,AU.
Thus, a rotationally supported disk can be sustained only in the magnetic dissipation region that has a size of $\sim5$\,AU (for details, see \S\ref{sec:diss-disk}).
It is expected that the disk quickly grows to a size of $\sim100$\,AU as the infalling envelope dissipates, and magnetic braking can no longer couple the disk to material in the envelope with a significant moment of inertia \citep{machida11b}. 
\citet{machida11b} showed that an exponential disk growth occurs and the disk can have a size of $\gtrsim 100$\,AU $\sim10^4-10^5$\,yr after protostar formation.

The small-sized disk seen in this simulation is consistent with recent observations around Class 0 protostars. When the protostar has a mass of $<0.1\msun$ (or in a very early stage of the gas accretion phase), the disk size is constrained to be $<10$\,AU \citep{yen17}. In other words, if the disk is not detected, the spatial resolution of the telescope yields an upper limit to the disk size. Our result is also consistent with recent numerical simulations using a central sink. 
\citet{zhao16,zhao18} showed that a rotationally supported disk does not appear when well-known grain properties are adopted.
Note that, in their study, a rotationally supported disk appears only when small-sized grains with size of $<0.1\,\mu$m are artificially removed from the MRN \citep{mathis77} dust size distribution. The disk does not appear with the standard MRN distribution that has minimum and maximum dust sizes of 0.005 and 0.25 $\mu$m, respectively.
This is because the small-size grains, which decrease the ionization degree and thus enhance the ambipolar diffusion and Ohmic dissipation rates, are not included.
Since they used a central sink of radius $2$\,AU, 
it is natural to not capture the disk in the gas accretion phase because the sink radius is comparable \citep{zhao16,zhao18} to or larger \citep{li11,li14b} than the actual disk size at these times.

\begin{figure}
\includegraphics[width=\columnwidth]{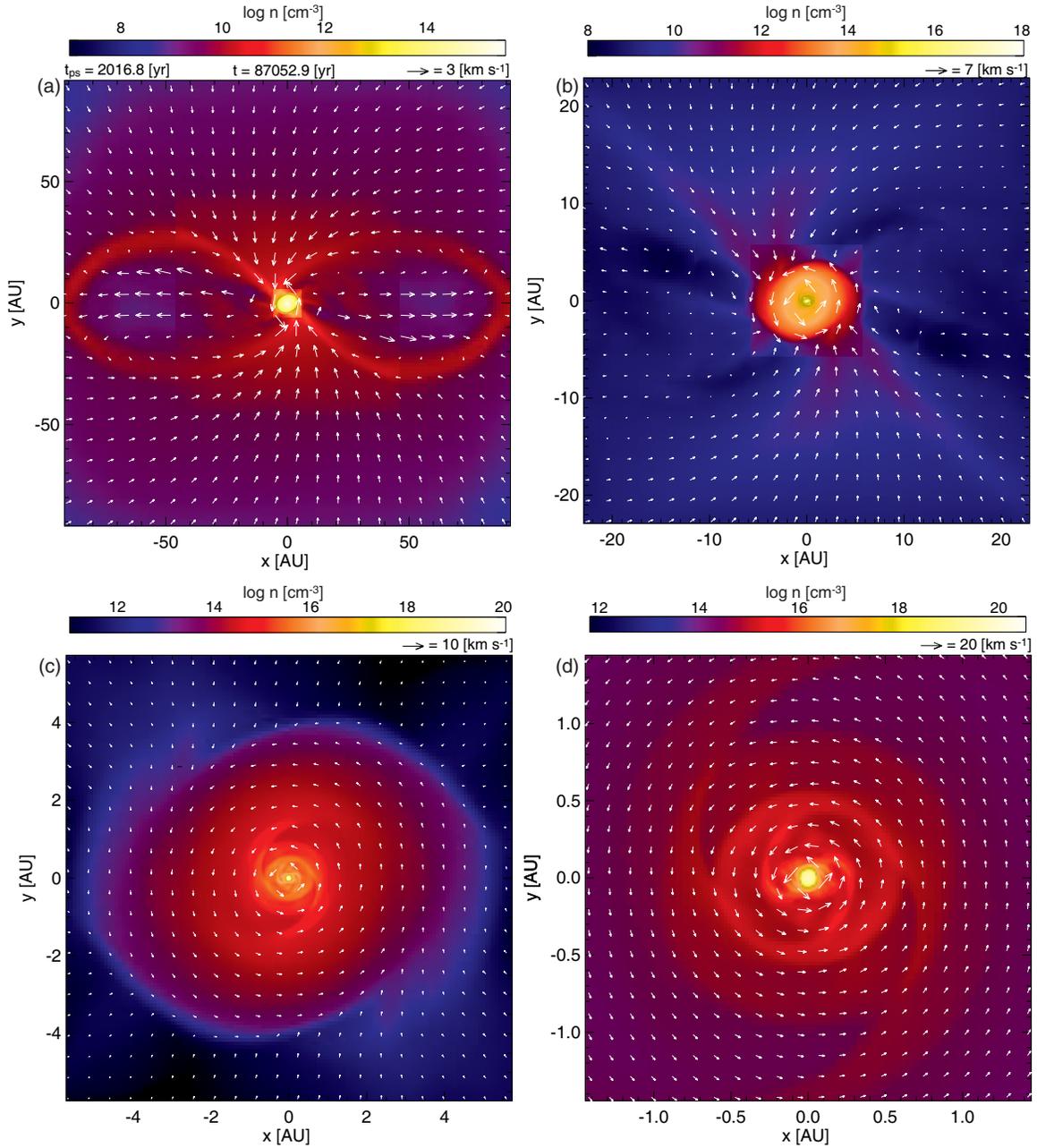}
\caption{
The density (color) and velocity (arrows) distributions on the $z=0$ plane at $t_{\rm ps}=2016.8$\,yr.
The spatial scale is different in each panel. 
 }
\label{fig:2}
\end{figure}

Figure~\ref{fig:2} shows the structures around the protostar at $t_{\rm ps}=2016.8$\,yr after protostar formation.
Two cavity-like structures along the $x$-axis can be seen on the large scale (Fig.~\ref{fig:2}{\it a}). 
The cavities appear only near the equatorial plane.  
This kind of cavity has been seen in some star formation simulations \citep[e.g.,][]{seifried11, krasnopolsky12,joos12, machida14, matsu17, zhao18} and is expected to be caused by the magnetic interchange instability.
We confirmed that the magnetic field in the cavity is much stronger than that around the cavity, which is a clear evidence for the interchange instability \citep{li96,tassis05}.
The cavity (or interchange instability) begins to develop just after the disk and protostar formation, while it becomes apparent at $t_{\rm ps}\sim 1000$\,yr.
Although the asymmetric structure appears due to the interchange instability, the gas can flow into the central region by avoiding the cavities as seen in Figures~\ref{fig:2}{\it a} and {\it b}.
Near the protostar, the density increases and the azimuthal velocity dominates the radial velocity (Figs.~\ref{fig:2}{\it c} and {\it d}). 
Figure~\ref{fig:2}{\it d} shows that spiral arms develop in the proximity of the protostar due to gravitational instability (for details, see \S\ref{sec:diss-disk}). 
Recent high-spatial resolution simulations have shown that a tiny rotationally supported disk forms around a very young protostar \citep{tomida13,machida14,tsukamoto15,vaytet18}. 
However, there is a controversy about the further evolution of the disk, i.e.,  whether or not such a tiny disk survives for a long duration and grows into a large-sized disk.
Our study shows that a rotationally-supported disk can be sustained for at least $\sim2000$\,yr after protostar formation. 
Further time integration is necessary to confirm whether the small-sized disk grows to a large-sized disk observed around older Class 0 or Class I objects.

\subsection{High-velocity Jet and Low-velocity Outflow} 
\label{sec:jet}
\begin{figure}
\includegraphics[width=\columnwidth]{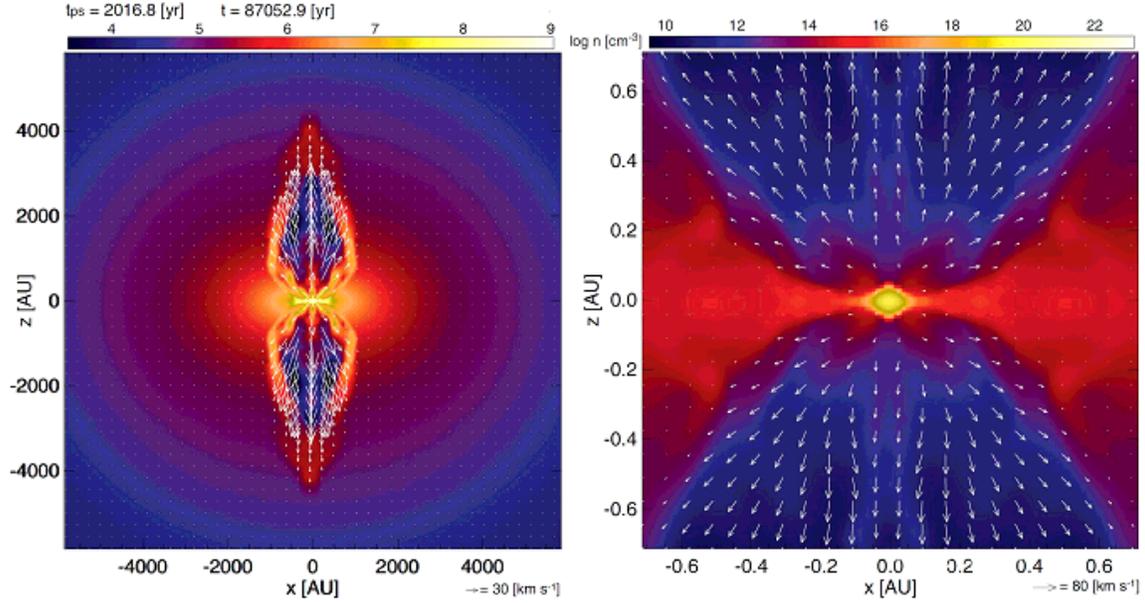}
\caption{
The whole structure of the outflow (left) and the region around the protostar (right) at the end of the simulation ($t_{\rm ps}=2016.8$\,yr).
The density (color) and velocity (arrows) distributions on the $y=0$ plane are plotted. 
}
\label{fig:3}
\end{figure}

We describe the mass ejection from the protostar and circumstellar disk in this subsection.
Figure~\ref{fig:3} shows the shape of the whole outflowing region (left) and the structure around the driving region of high-velocity component (right) at the end of the simulation. 
The root of the high-velocity flow is located near the surface of the protostar, which has a size of $<0.1$\,AU (Fig.~\ref{fig:3} right). 
The high-velocity flow gradually accelerates toward the $z$-direction and the flow velocity reaches $\sim100\km$.
The outflowing region extends up to $\sim4000$\,AU during $\sim2000$\,yr after protostar formation (Fig.~\ref{fig:3} left).
Thus, we can calculate the outflow propagation until the flow reaches a scale of $\sim1000$\,AU, while resolving its driving region that has a scale of $\lesssim 0.1$\,AU. 
The difference in the scales between the flow-driving region ($<0.1$\,AU) and the whole outflowing region ($>10^3$\,AU) is $>10^4$.
The outflow size of $>1000$\,AU is sufficient to directly compare the simulation with observations.
However, it should be noted that the spatial resolution around  the head of the outflow may not be  sufficient. 
As the outflow evolves, the low-resolution grids cover the region around the head of the outflow. 
Thus, we cannot resolve the  whole region of the evolved outflow with high spatial resolution. 
In addition, we have no clear criteria about the necessary condition for spatial resolution of the outflow.
An investigation of this is a subject for a future study, but beyond the scope of the present study.

\begin{figure}
\includegraphics[width=\columnwidth]{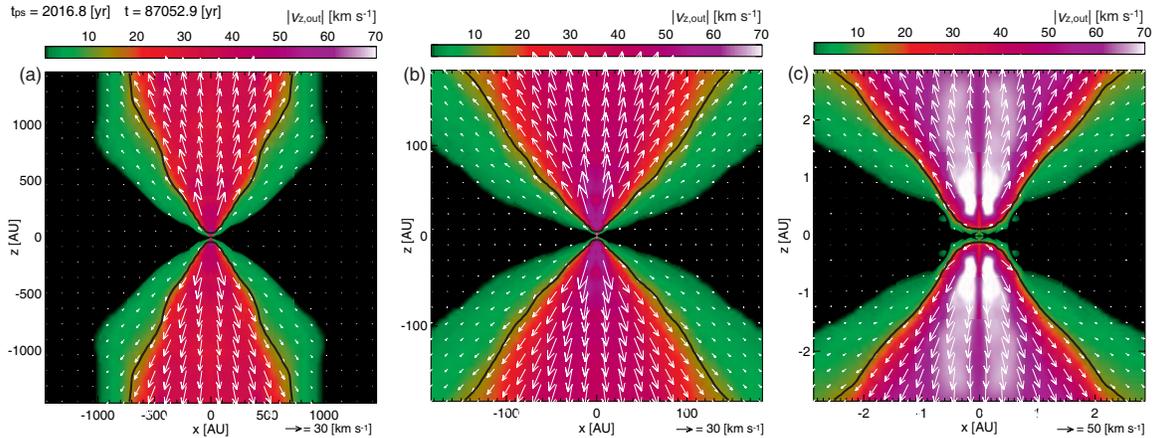}
\caption{
The absolute value of the $z$-component of the outflow velocity $\vert v_z \vert$ is plotted in color in each panel. 
The spatial scale differs in each panel. The arrows indicate the velocity vector. The iso-velocity of $\vert v_z \vert = 10\km$ is plotted by the black contour inside which the outflow velocity exceeds $\vert v_z \vert >10\km$.
}
\label{fig:4}
\end{figure}

Figure~\ref{fig:4} plots the velocity distribution inside the outflowing region in different spatial scales at the end of the simulation, in which the outflow velocity is represented by color. 
In the large (Fig.~\ref{fig:4}{\it a}) and middle (Fig.~\ref{fig:4}{\it b}) scales, the low-velocity component of $\lesssim 10\km$ (colored by green) seems to be separated from the high velocity component of $\gtrsim 10\km$ (colored by red).
For convenience, we call the low-velocity component `the low-velocity outflow (or the outflow, $v_{\rm out} \lesssim 10\km$)' and the high-velocity component `the high-velocity jet (or the jet, $v_{\rm out} \gtrsim  10\km$)', respectively. 

The two component flows (the low-velocity outflow and high-velocity jet) are often seen in MHD simulations of collapsing star-forming cloud cores \citep[e.g.,][]{tomisaka02,banerjee06,machida06,tomida13,bate14}. 
Since the driving mechanism has been well explained in many past studies, we simply summarize it.
The flows are driven by the magnetic effects and their driving mechanisms are the magneto-centrifugally driven \citep{blandford82} and magnetic pressure driven \citep[or magnetic tower;][]{uchida85,lynden-Bell03} mechanisms, respectively.
Past studies have shown that the low-velocity wide-angle flow (or the outflow) is driven by the first core before protostar formation (or in the gas collapsing phase) and the outer edge of the disk after protostar formation (or in the gas accretion phase).
The magnetic field is well coupled with the neutral gas in the outer edge of the first core, or outer disk region, where the (low-velocity) outflow is driven.
On the other hand, the magnetic field is not coupled with the neutral gas inside the disk because the dust grains absorb the charged particles and the ionization degree becomes extremely low. 
As a result, no magnetic driven flow appears inside the disk. 
However, near the protostar (or disk inner edge), the high gas temperature enhances the ionization degree, and the magnetic field is again coupled with the neutral gas.
Thus, another magnetic driven flow (i.e., the high-velocity jet) appears around the disk inner edge. 
The flow speed roughly corresponds to the Keplerian velocity at its driving region \citep{machida08b}.
Thus, two distinct flows (low-velocity outflow and high-velocity jet) appear, in which one flow is driven by each magnetically active region \citep{tomida13,machida14}.

We can additionally see a higher velocity component ($\gtrsim70\km$) on a very small scale (Fig.~\ref{fig:4}{\it c}). 
This component appears in the region just above and below the protostar, and is directly related to the episodic accretion as described in \citet{machida14}. 
Although we could define this component as 'the third outflow,' we include its description into `the high-velocity jet' in order to avoid complicated definitions. 
Although two or three distinct flows were seen in some previous MHD simulations \citep{tomisaka02,banerjee06,machida06,tomida13}, the calculations were stopped before the flows grew sufficiently.

\begin{figure}
\includegraphics[width=\columnwidth]{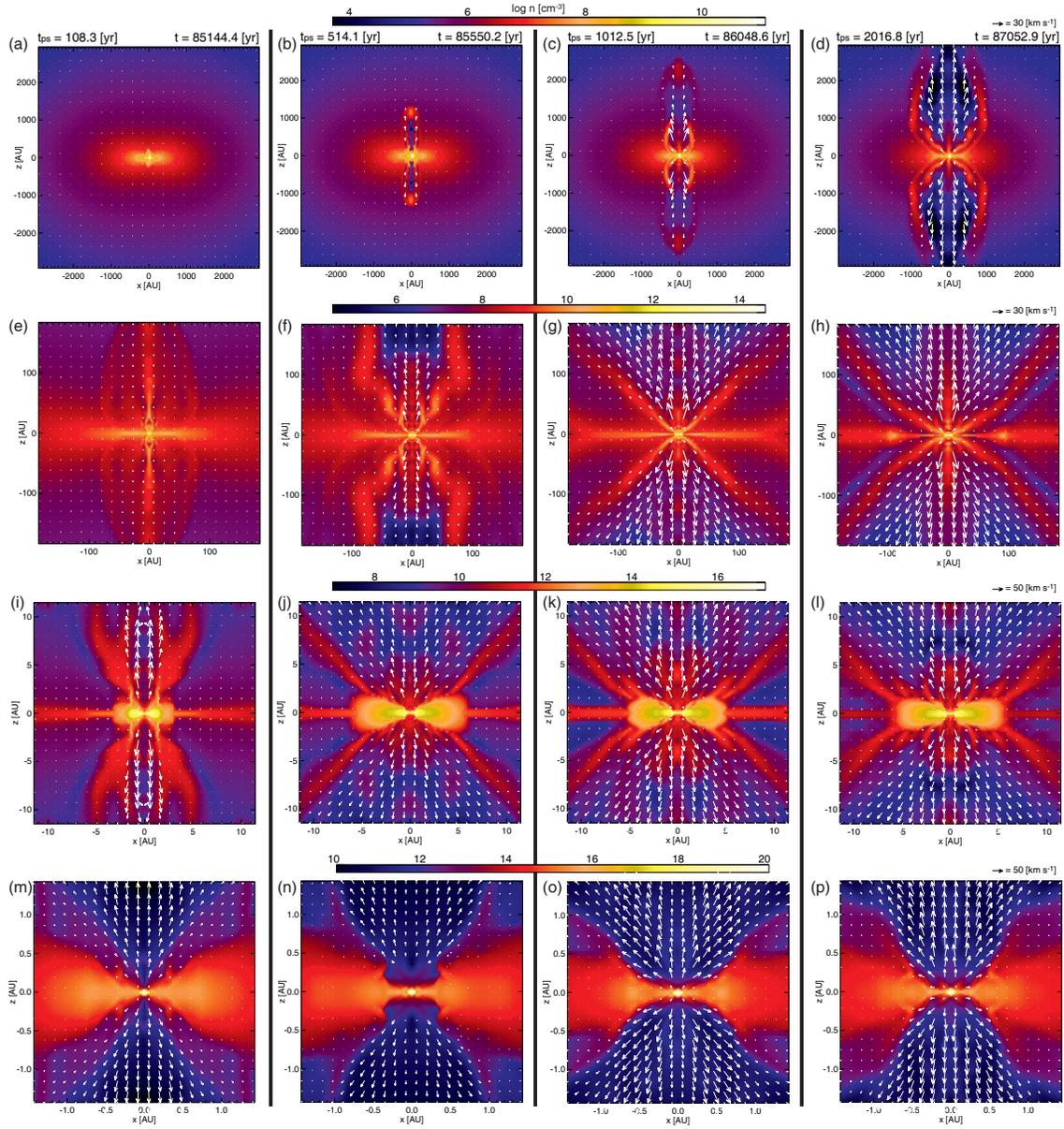}
\caption{
Time evolution of the low-velocity outflow and high-velocity jet. 
The density (color) and velocity (arrows) distributions on the $y=0$ plane are plotted. 
The epoch is the same in each column, while the spatial scale is the same in each line. 
The elapsed time $t_{\rm ps}$ after protostar formation and total time $t$ after the initial cloud begins to collapse are described in the top panel of each column.  
(A movie is available in the on-line journal.)
}
\label{fig:5}
\end{figure}

Figure~\ref{fig:5} shows the time evolution of the low-velocity outflow and the high-velocity jet for about 2000\,yr after protostar formation. The spatial scale differs in each row.  
Following protostar formation (the first column of Fig.~\ref{fig:5}), the high-velocity jet is driven from around the protostar (Fig.~\ref{fig:5}{\it i} and {\it m}), while it does not catch up with the head of the low-velocity outflow (Fig.~\ref{fig:5}{\it a} and {\it b}). 
The low-velocity outflow appears after the first core formation and before the protostar formation, while the high-velocity jet appears just after protostar formation.
Thus, the outflow precedes the jet in early gas accretion phase.
Then, the jet catches up with the outflow in $t_{\rm ps}\lesssim 300$\,yr as seen in Fig.~\ref{fig:5}({\it b}) -- ({\it d}).
Thereafter, the jet pushes out the outflow from the inside to the outside and creates a low-density cavity (e.g., Fig.~\ref{fig:5}{\it f} and {\it d}).
Although the jet speed gradually increases with time, the structure around the protostar does not change significantly (Fig.~\ref{fig:5}{\it m} -- {\it p}).
On the other hand, the outflow opening angle widens with time on a large scale, as seen in Figure~\ref{fig:5}{\it i}--{\it l} and Figure~\ref{fig:5}{\it e} -- {\it h}.
The observations also show that the outflow opening angle widens as the protostellar system evolves.
In addition, a capital `X'-like structure (Fig.~\ref{fig:5}{\it g} and {\it h}) and cavity and shell-like structure  seen in many observations \citep{velusamy07,velusamy14} appear in the simulation. 

In Figure~\ref{fig:5}({\it d}), we can also see a horn-like structure near the equatorial plane and inside the cavity. 
The horn is created by the high-velocity jet, while the outer shell that shapes the large-scale cavity is created by the low-velocity outflow
\footnote{
We define a horn (or horn-like structure) as the object created by the high-velocity jet and located inside the large-scale cavity that is made by the low-velocity outflow. Some horns can be seen in Figure~\ref{fig:5}{\it l}.  
In addition, we use the term `X-like structure' when discussing their morphology and comparing them with observations.
}. 
Although the inner horn may merge with the outer shell in the future, its existence is a proof of two outflow components.  

\begin{figure}
\includegraphics[width=\columnwidth]{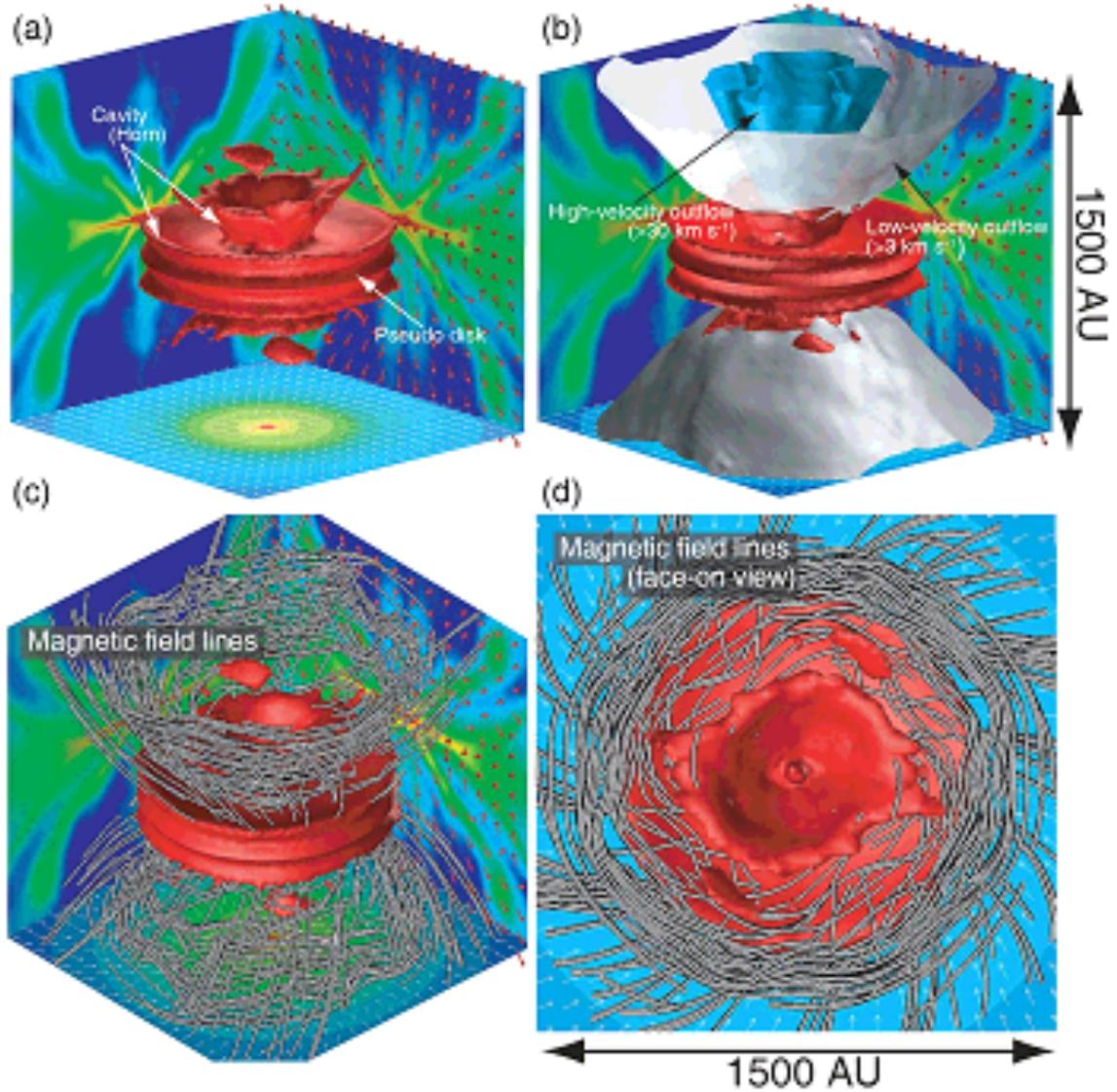}
\caption{
Three dimensional view made from data at the end of the simulation.
(a) Iso-density surface of $n=10^8\cm$ is plotted by the red surface. 
The density and velocity distributions on $x=0$, $y=0$ and $z=0$ plane are projected on each wall surface.
(b) Iso-velocity surface of $v_z=30$\,km\,s$^{-1}$ (blue) and  $v_z=3$\,km\,s$^{-1}$ (gray) are added to panel (a).
(c) Magnetic field lines are added to panel (c).
(d) Same as in panel (c), but the viewing angle differs.
}
\label{fig:6}
\end{figure}

To clearly see the horn and outflow structures, three dimensional views are presented in Figure~\ref{fig:6}.
The box scale of Figure~\ref{fig:6} is 1500\,AU, and the iso-density surface of $n=10^8\cm$ is plotted by a red surface that traces the inner (or horn) and outer cavities above and below the pseudo disk.
In addition, the density and velocity distributions on $x=0$, $y=0$ and $z=0$ are projected on each wall surface. 
Figure~\ref{fig:6}{\it b} is the same figure as Figure~\ref{fig:6}{\it a}, but the iso-velocity surface of the low-velocity outflow (gray, $v_z=3\km$) and the high-velocity jet (blue, $v_z=30\km$) are added.
From these figures, we can confirm that the inner horn is created by the high-velocity jet and the outer cavity is created by the low-velocity outflow. 
Thus, two distinct flows (outflow and jet) produce the inner and outer cavities as seen in Figure~\ref{fig:6}{\it a} and \ref{fig:5}{\it d}. 
The magnetic field lines are plotted in Figure~\ref{fig:6}{\it c} and {\it d}, in which only the viewing angle (face-on view) is changed in Figure~\ref{fig:6}{\it d}.
These figures indicate that the magnetic field lines are strongly twisted only inside the outflowing region and they are loosely coiled outside the outflow. 
Radially directed magnetic field lines outside the low-velocity outflow have been seen in recent observations \citep[e.g.,][]{alves18}.

\begin{figure}
\includegraphics[width=\columnwidth]{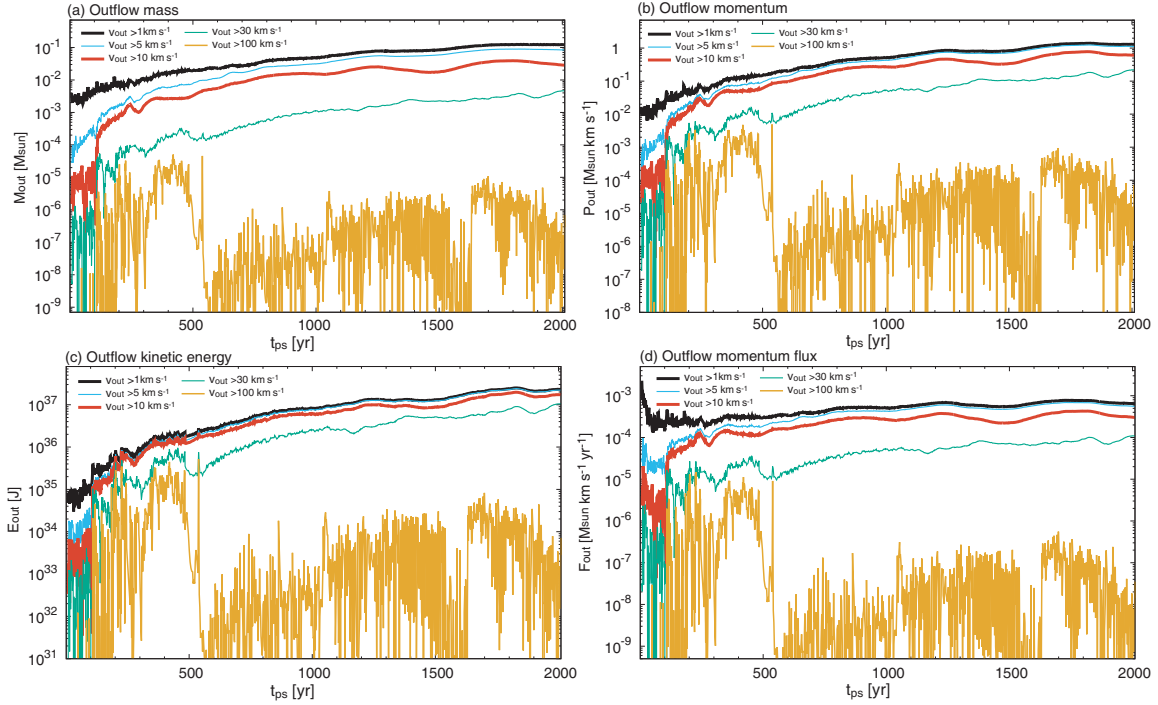}
\caption{
Outflow physical quantities: (a) outflow mass, (b) outflow momentum, (c)  outflow kinetic energy, and (d) outflow momentum flux with different velocity thresholds, are plotted against the time elapsed after protostar formation.
}
\label{fig:7}
\end{figure}

Next, we describe the outflow physical quantities. 
Figure~\ref{fig:7} shows the outflow mass (Fig.~\ref{fig:7}{\it a}), momentum (Fig.~\ref{fig:7}{\it b}), energy (Fig.~\ref{fig:7}{\it c}) and momentum flux (Fig.~\ref{fig:7}{\it d}) versus the elapsed time after protostar formation, in which each quantity is estimated with different threshold velocities.
These quantities are calculated as 
\begin{equation}
M_{\rm out} = \int_{v_{\rm out} > v_{\rm cri}} \rho \, dV,
\end{equation}
\begin{equation}
P_{\rm out} = \int_{v_{\rm out} > v_{\rm cri}} \rho \, v_{\rm out} dV,
\end{equation}
\begin{equation}
E_{\rm out} = \frac{1}{2} \int_{v_{\rm out} > v_{\rm cri}} \rho \, v_{\rm out}^2 dV,
\end{equation}
\begin{equation}
F_{\rm out} = \frac{\int_{v_{\rm out} > v_{\rm cri}} \rho \, v_{\rm out} \, dV}{t_{\rm ps}},
\end{equation}
where $v_{\rm out}$ is the outflow velocity and the threshold velocities are adopted as $v_{\rm cri}=1$, 5, 10, 30, 100$\km$. 
The outflow physical quantities shown in Figure~\ref{fig:7} are in good agreement with observations \citep{wu04}.
Figure~\ref{fig:7} also indicates that the low-velocity component ($v_{\rm out} \le 1$, 5, 10\,$\km$) is more dominant than the high-velocity component ($v_{\rm out} \ge 30$, 100$\, \km$) in each of the above quantities.
In Paper I, the propagation of the high-velocity jet was calculated for only $\sim200$\,yr. 
Since the low-velocity outflow, which is originally driven by the first core, appears much before the appearance of the high-velocity jet, we need to calculate the jet propagation for a much longer time, as done in this paper, in order to evaluate the effect of the jet on the star formation process. 
Figure~\ref{fig:7} shows that the contribution of the jet is actually not dominant.
Our results indicate that the low-velocity outflow driven by the outer disk edge contributes relatively more than the jet to the overall star formation process. 
As described in \S\ref{sec:mass}, the low-velocity outflow is responsible for the ejection of a large mass fraction of the host cloud core, and thereby may affect the determinant of the star formation efficiency.
Since the low-velocity outflow is driven at the disk outer edge, it has a large radius and wide opening angle by definition. 
On the other hand, the high-velocity jet cannot be ignored because it still donates significant momentum to the low-velocity outflow and to the infalling matter.

\subsection{Mass of Each Object}
\label{sec:mass}
This subsection focuses on the masses of the protostar, disk, and outflow.
We did not use a sink cell.
Thus, we need to define each object (the protostar, disk and outflow) in order to estimate their mass, because there is no clear boundary between the protostar and its surrounding disk (or outflow).  

We define the protostar as the region having a number density of $n>10^{18}\cm$ and a radial velocity of  $v_r < 1\km$. 
Note that the latter condition ($v_r < 1\km$) was imposed in order to exclude the outflow component.
We integrated the mass of each cell satisfying these conditions, and estimated the protostellar mass $M_{\rm ps}$.

The disk is defined as the region satisfying the following conditions:
(1) the number density is in the range of  $10^7 \cm < n < 10^{18}\cm$; 
(2) the azimuthal velocity is greater than the radial velocity, i.e., $v_\phi > v_{\rm r,c}$, where $v_{\rm r,c}$ is the radial velocity in a cylindrical coordinate system; 
and 
(3) the azimuthal velocity is greater than 90\% of the Keplerian velocity, i.e., $v_\phi > 0.9\, v_{\rm Kep}$, where the Keplerian velocity is $v_{\rm Kep}=(G\,M_{\rm ps}/r)^{1/2}$
\footnote{  
The condition (3) does not effectively  play a role to identify the disk in the very early gas accretion phase, because the disk mass is comparable to or exceeds the protostellar mass. The condition (2) is able  to identify the disk  in such an early phase \citep{machida10a}, while the condition (3) is necessary to determine it in the later phase.
}.
We integrated the mass of each cell satisfying the conditions (1) -- (3) in order to estimate the disk mass $M_{\rm disk}$.
The outflow mass $M_{\rm out}$ is calculated according to the procedure described in \S\ref{sec:jet}, in which we adopt $v_{\rm cri}=1\km$.
We note that these criteria were determined to be the most plausible after comparing various possible criteria with estimates made by eye.

\begin{figure}
\includegraphics[width=\columnwidth]{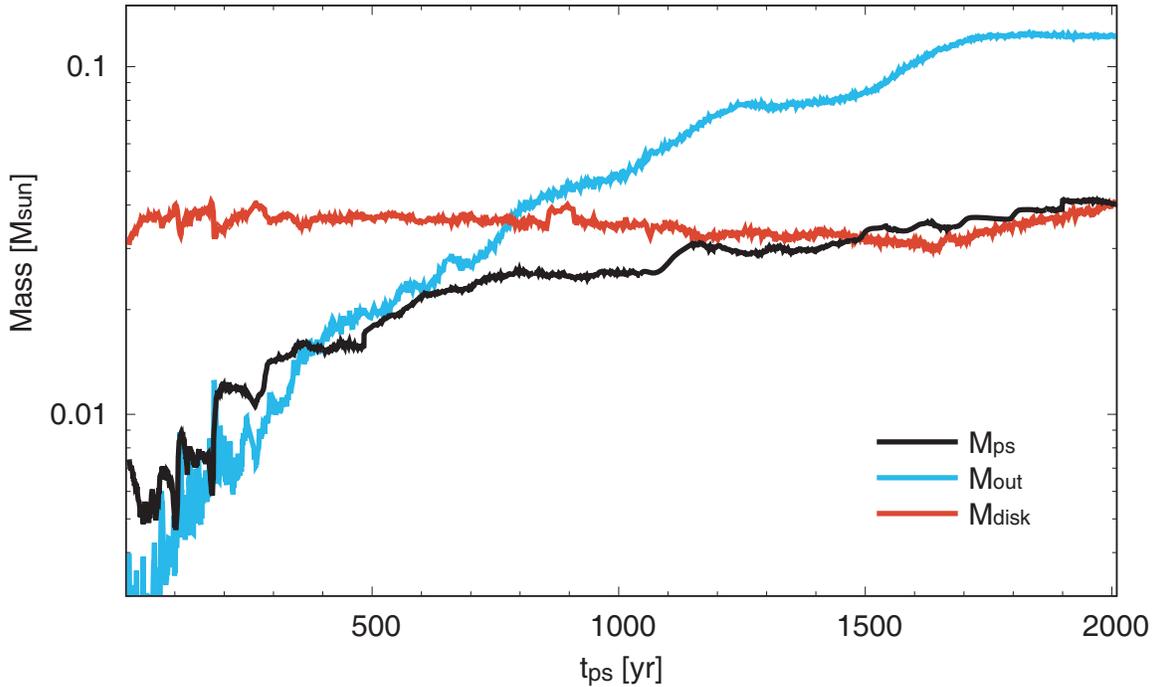}
\caption{
Masses of the protostar $M_{\rm ps}$, circumstellar disk $M_{\rm disk}$, and outflow $M_{\rm out}$ against the time elapsed after protostar formation.
}
\label{fig:8}
\end{figure}

Figure~\ref{fig:8} plots the mass of protostar, disk, and outflow against the elapsed time after protostar formation. The protostar and outflow increase their mass with time,  while the disk mass does not significantly increase for $\sim2000$\,yr. 
Note that, in Figure~\ref{fig:8}, a temporal decrease of the protostellar mass is due to the slight expansion of the protostellar surface during which a part of the protostellar surface temporarily decreases to  $<10^{18}\cm$ and is not identified as the protostar. 
During $\sim1000$ yr after protostar formation (i.e., $t_{\rm ps}\lesssim 1000$\,yr), the disk is more massive than the protostar, because a large part of the first core remnant, which has a mass $\gtrsim0.01\msun$, becomes the circumstellar disk. Conversely, the protostar has a mass $\simeq 0.001\msun$ at its formation epoch ($t_{\rm ps}\simeq0$\,yr). 
As seen in Figure~\ref{fig:2}{\it d}, gravitational instability tends to occur when the disk mass exceeds the protostellar mass \citep{toomre64} and a significant fraction of the disk gas quickly falls onto the protostar. 
Therefore, the protostar increases its mass in a short duration. 
The mass evolution of protostar and disk during the early mass accretion phase is well explained in \citet{inutsuka10}.

Figure~\ref{fig:8} also indicates that the rate of increase of the outflow mass is greater than that of the protostellar mass. 
As described in \S\ref{sec:jet}, the outflow opening angle gradually widens with time.
Thus, a large fraction of the infalling gas is swept up by the (original) outflow which is directly driven from the disk. 
At $t_{\rm ps}=2024$\,yr, which corresponds to the end of the simulation, the protostar, disk, and outflow have masses of $M_{\rm ps}=0.04\msun$,  $M_{\rm disk}=0.041 \msun$, and $M_{\rm out}=0.12\msun$, respectively. 
Thus, the outflow mass ($M_{\rm out}=0.12\msun$) is greater than the protostellar mass ($M_{\rm ps}=0.04\msun$), and over half of the infalling material is blown away by the outflow. 
Therefore, we infer that the outflow can have an impact in determining the star formation efficiency, if the high outflow activity lasts until the end of the gas accretion phase \citep[see also][]{machida13}.

\begin{figure}
\includegraphics[width=\columnwidth]{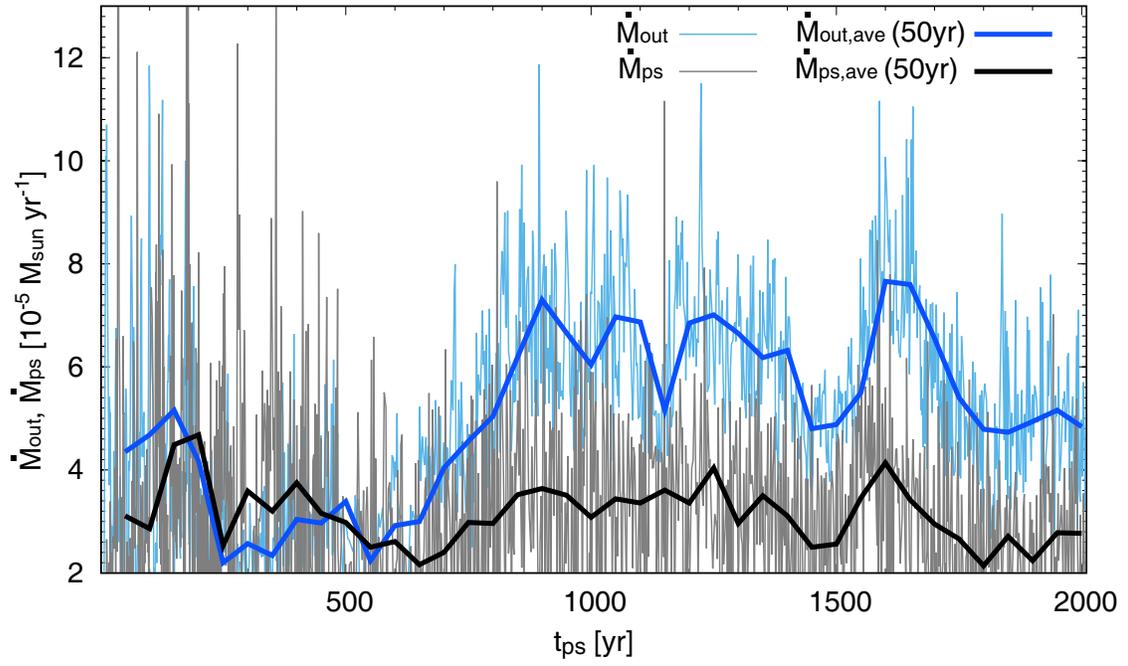}
\caption{ 
Instantaneous mass accretion (thin blue) and outflow (thin black) rates are plotted against the elapsed time after protostar formation. 
The mass accretion (thick blue) and outflow (thick black) rates averaged over 50\,yr are also plotted.  
}
\label{fig:9}
\end{figure}

The masses of the protostar and outflow are also the time integrated values of the mass accretion and outflow rates, respectively. For completeness, we also estimate the instantaneous mass accretion and outflow rates in the following manner.
Both quantities are calculated on the surface of a grid \citep{tomisaka02,matsushita17}.  
The mass accretion rate is calculated as 
\begin{equation}
\dot{M}_{\rm acc} (l_{\rm acc}) = \int_{\rm surface \, of \, level\, {\it l=l_{\rm acc}} \ grid} \rho \, \vect{v} \cdot \vect{n} (<0) \, dS,
\end{equation}
where $\vect{n}$ is the normal vector of each surface and we executed the integration only when $\vect{v} \cdot \vect{n}<0$ (i.e., inflow).
The mass outflow rate is calculated as  
\begin{equation}
\dot{M}_{\rm out} (l_{\rm out}) = \int_{\rm surface \, of \, level\, {\it l=l_{\rm out}} \ grid} \rho \, \vect{v} \cdot \vect{n} (>0) \, dS,
\end{equation}
where we executed the integration only when $\vect{v} \cdot \vect{n}>0$ (i.e., outflow).
In the equations, we adopted $l_{\rm acc}=21$ and $l_{\rm out}=14$, respectively. 
To estimate the mass accretion, we adopted the finest grid ($l=21$, $L_{\rm box}[l=21]=0.36$\,AU) which just covers the protostar (Fig.~\ref{fig:3} right).
On the other hand, we adopted $l=14$ ($L_{14}=46$\,AU) to estimate the outflow rate because the outflow driving region extended to $\sim10$\,AU (Fig.~\ref{fig:5}). 
With this grid level ($l=14$), we can roughly trace the outflow directly driven by the disk, while we may fail to trace the outflowing gas swept up by this outflow.
Thus, the total outflow rate may be underestimated in this analysis.
Figure~\ref{fig:9} plots the estimated instantaneous (thin lines) and averaged (thick lines) mass accretion and outflow rates against the elapsed time after protostar formation.
It shows that the outflow rate is greater than mass accretion rate, especially for $t_{\rm ps} \gtrsim 500$\,yr. 
This indicates that the gravitational energy released in the accretion process can be efficiently converted to the kinetic energy of the outflow through the magnetic effect. 

\begin{figure}
\includegraphics[width=\columnwidth,pagebox=artbox]{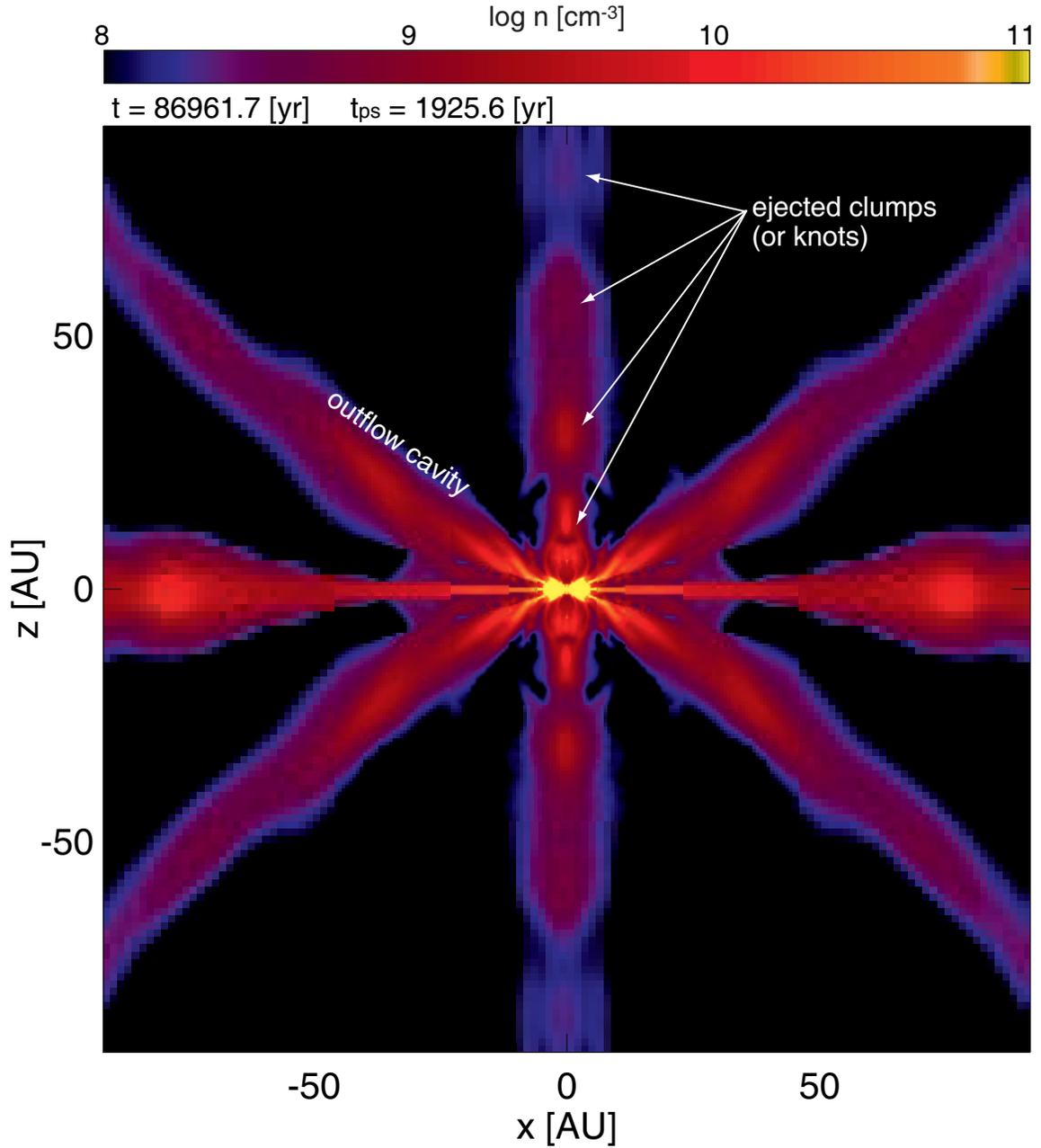}
\caption{
The density distribution on the $y=0$ plane at $t_{\rm ps}=1925.6$\,yr. The ejected clumps are  indicated by arrows and the outflow cavity is labelled.
}
\label{fig:55}
\end{figure}

In Figure~\ref{fig:9}, in addition to the mass accretion rate (see \S\ref{sec:disk}), the mass ejection rate shows a high time variability, which implies an episodic mass ejection. 
The density distribution on the $y=0$ plane is plotted in Figure~\ref{fig:55}, in which the density range is adjusted to stress the clumps ejected from the central region associated with the high-velocity jet. 
In addition to the cavity created by the low-velocity outflow, the presence of several knots is confirmed in the figure. 
A recent observation also showed a clumpy structure of the high-velocity jet in the very young protostellar system \citep{matsushita19}. 
As described in \S\ref{sec:disk}, the gravitational instability induces the intermittent mass accretion, which then drives the episodic jet that is associated with several knots.

\section{Discussion}
\label{sec:discussion}
\subsection{Formation of a Rotationally Supported Disk}
\label{sec:diss-disk}
As described in \S\ref{sec:disk}, there exist in theoretical studies some controversial issues about disk formation, i.e., when and how the circumstellar (or rotationally supported) disk forms \citep{mellon08,li11,machida11b,machida14b}. 
Some researchers pointed out that magnetic braking suppresses disk formation in the early phase of the star formation, so that the disk is expected to be formed in the later phase of the star formation \citep{li14}. 
However, in the ALMA era, many circumstellar disks (or Keplerian disks) have been observed around Class 0 objects \citep{yen17}. 
Thus, the observations indicate that a circumstellar disk forms in the early gas accretion phase.

There are two types of theoretical (or simulation) studies about circumstellar disk formation that include non-ideal MHD effects.
A sink simulation is one type, in which the sink method is used to accelerate the time integration, and the cloud evolution is calculated for $\gtrsim 10^3-10^4$\,yr after the sink creation (or protostar formation).
In this type of study, the protostar and the circumstellar region ($\lesssim 1-10$\,AU) cannot be resolved. 
In addition, the disk formation condition strongly depends on the numerical settings such as the spatial resolution and sink treatment \citep{machida14}. 
A circumstellar disk does not appear at these times when using a large sink radius, while it does appear if using a sufficiently small sink radius.
The second type is a simulation study without the sink method, in which the protostar is resolved. 
A rotationally-supported disk appears in all of these types of simulations \citep[e.g.,][]{dapp10,dapp12,tomida13,machida14,tsukamoto15,vaytet18,wurster18b}. 
However, since the calculation time step becomes extremely short, the cloud evolution cannot be calculated for $\gtrsim 1-10$\,yr after protostar formation. 
\citet{machida14} calculated the cloud evolution for only $>100$\,yr after protostar formation while resolving the protostar. In that simulation a disk with a size of $\lesssim 2$\,AU was maintained. Our current simulations are better able to connect simulations and observations because we could calculate the cloud evolution for $\sim2000$\,yr after protostar formation. For even more complete comparisons, future simulations will need to calculate the disk evolution for an even much longer duration, while resolving the protostar. 

\begin{figure}
\includegraphics[width=\columnwidth]{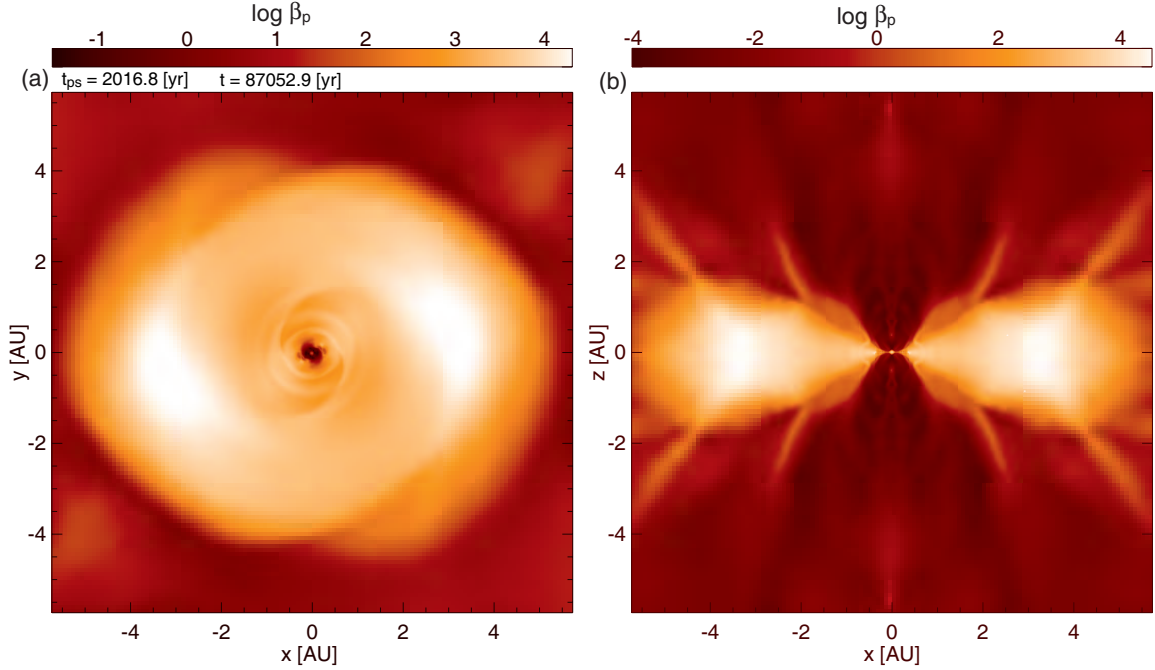}
\caption{
Plasma beta $\beta_{\rm p}$ (color) on the $z=0$ (left) and $y=0$ (right) plane at the end of the simulation.
}
\label{fig:10}
\end{figure}

As described in \S\ref{sec:disk}, the disk is sustained for about $2000$\,yr after protostar formation.
\citet{mellon08} claimed that magnetic braking suppresses disk formation  in the ideal MHD limit, while \citet{machida11b} pointed out that disk formation is possible in the magnetically inactive region where the magnetic field dissipates by Ohmic dissipation and ambipolar diffusion \citep[see also][]{dapp10,dapp12}. 
This study showed that a rotationally-supported disk can be sustained for at least  $2000$\,yr after its formation, although the disk size is limited to $\sim$\,AU scales. 
To confirm the effect of the magnetic field and its dissipation, the plasma beta, i.e., $\beta_{\rm p}\equiv (8\pi P)/B^2$, in which $P$ is the thermal pressure, is plotted at the end of the simulation in Figure~\ref{fig:10}.
The left panel shows that $\beta_{\rm p}$ has a peak around $r\simeq 3-4$\,AU and a high beta region ($\beta_{\rm p}>10^2$) extends from $0.1$\,AU to $\sim4$\,AU. 
The high beta region corresponds to the rotationally-supported disk (compare Fig.~\ref{fig:10} left side with Fig.~\ref{fig:2}{\it c}).
The Ohmic dissipation becomes effective in the density range of $10^{11}\cm < n < 10^{15}\cm$ \citep{nakano02}, so the rotationally-supported disk can exist only in a high-density region where the magnetic field is not coupled with the neutral gas and magnetic braking therefore becomes ineffective.
It is expected that the disk size gradually increases in a later stage because the massive infalling envelope, which brakes the disk by accepting its angular momentum, gradually dissipates with time \citep{machida11b}. 
The growth time scale of the disk is estimated to be $\sim10^4-10^5$\,yr in \citet{machida11b}.
Thus, a further time integration is necessary to connect this study to simulations with sink cells.  

We can also confirm the structure of the outflowing region from the distribution of the plasma beta.
Figure~\ref{fig:10}{\it b} shows the plasma beta on a cut through the $y=0$ plane.
The figure indicates that the low-beta region well traces the outflowing region (see Fig.~\ref{fig:3} -- Fig.~\ref{fig:5}). 
The plasma beta is in the range of $\beta_{\rm p}<10^{-2}$ in the outflowing region.
Thus, the magnetic pressure is over 100 times larger than the thermal pressure, which indicates that the magnetic field greatly contributes to the outflow driving. 
Figure~\ref{fig:5} shows two horn-like structures above and below the disk, in which the inner horn is created by the high-velocity jet driven near the protostar, while the outer horn is created by the low-velocity outflow driven by the disk outer region. 
The horn-like structures are clearly confirmed also in Figure~\ref{fig:10}{\it b}.

\begin{figure}
\includegraphics[width=\columnwidth]{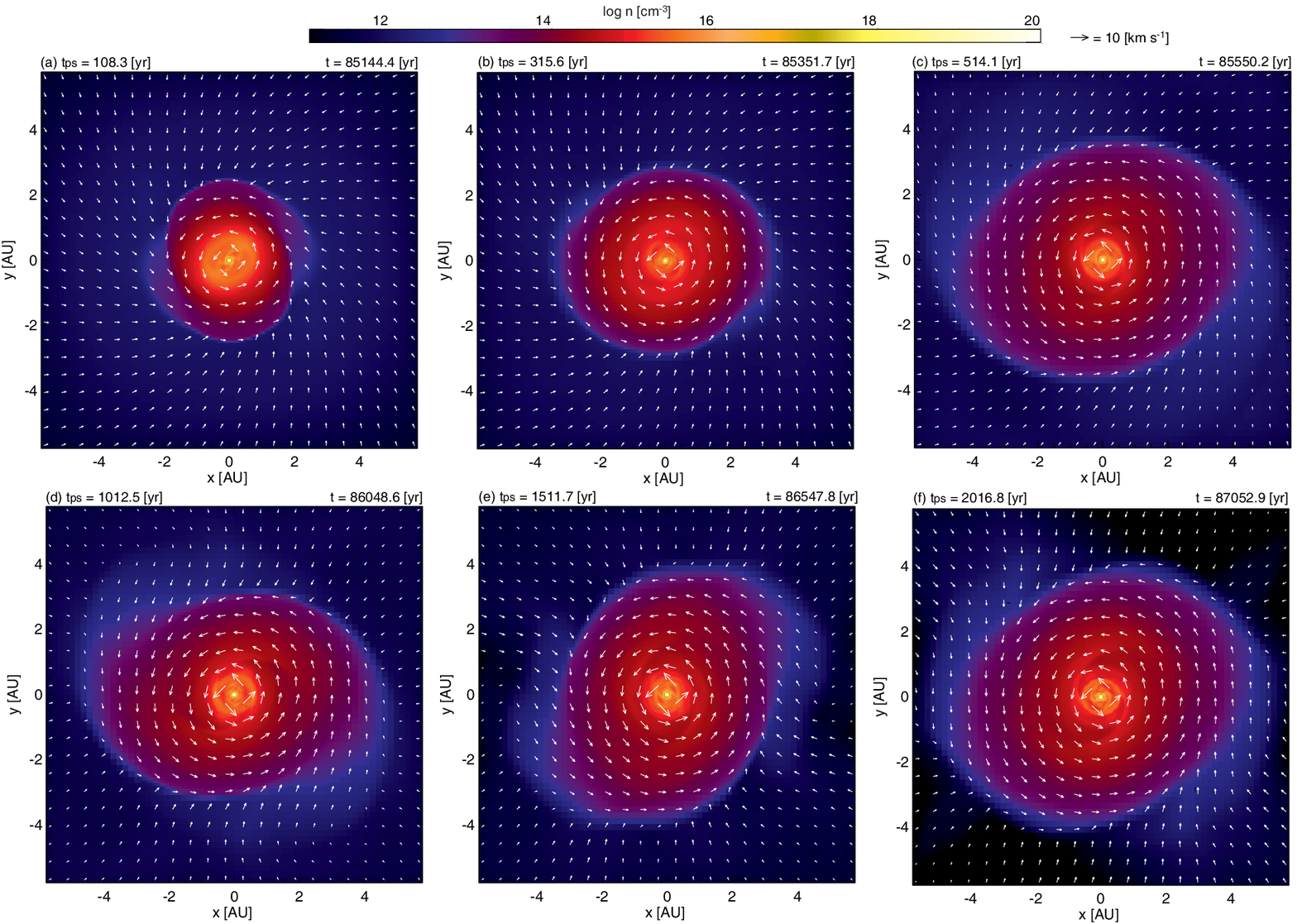}
\caption{
Time evolution of the circumstellar disk. The density (color) and velocity (arrows) distributions on the $z=0$ plane are plotted in each panel. 
The elapsed time $t_{\rm ps}$ after the protostar formation and the total time $t$ after the initial cloud begins to collapse are also described in each panel.  
The spatial scale is the same in the panels.
}
\label{fig:1}
\end{figure}

\begin{figure}
\includegraphics[width=\columnwidth]{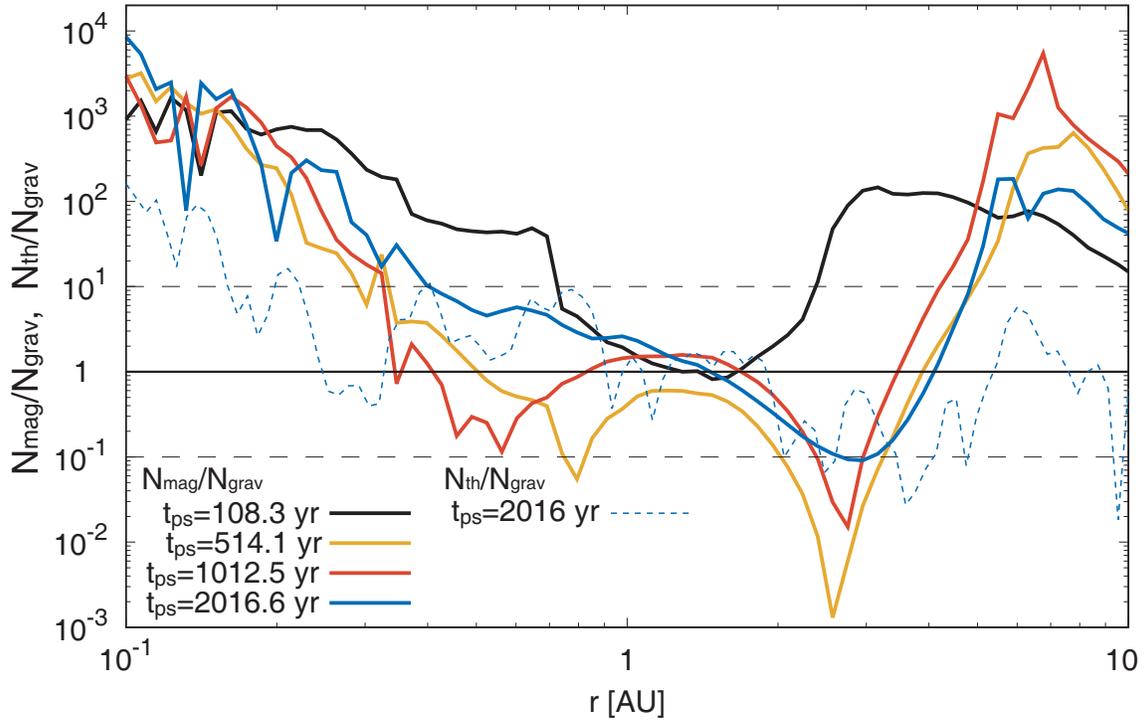}
\caption{
The ratio of magnetic to gravitational torques $N_{\rm mag}/N_{\rm grav}$ on the equatorial plane at 
four different epochs versus the radial distance from the protostar. 
The ratio of thermal to gravitational torques $N_{\rm th}/N_{\rm grav}$ at the end of the calculation is also plotted. 
}
\label{fig:11}
\end{figure}

In the magnetically active region where the magnetic field is well coupled with the neutral gas, the angular momentum is effectively transferred by the magnetic braking and the outflow.
On the other hand, in the magnetically inactive region (inside the disk), the self-gravity contributes to the angular momentum transfer. 
To confirm which effect (Lorentz force or gravity) is dominant in the angular momentum transfer, we estimated the magnetic and gravitational torques on the equatorial plane at four different epochs. 
The magnetic torque is estimated as 
\begin{equation}
N_{\rm mag}(r) = \int_r^{r+\Delta r} \vect{r} \times \left[ \frac{1}{4\pi}(\nabla \times \vect{B}) \times \vect{B} \right]\, dV, 
\end{equation} 
where $\Delta r$ is set to be comparable to the cell width $h(l)$.
The gravitational torque is estimated as
\begin{equation}
N_{\rm grav}(r) = \int_r^{r+\Delta r} \rho \, (\vect{r} \times \vect{g})\, dV, 
\end{equation} 
where $\vect{g}$ is the gravity (self-gravity and gravity of the protostar) at each cell. 
In addition, we also estimated the gas pressure (or thermal) torque as
\begin{eqnarray}
N_{\rm th}(r) = \int_r^{r+\Delta r} (\vect{r} \times  \nabla P)  \, dV,
\end{eqnarray}
where $P$ is the gas pressure. 
Figure~\ref{fig:11} shows the ratio of magnetic to gravitational torque plotted against the distance from the protostar, in which the ratio of thermal to gravitational torque only at the end of the simulation is also shown.   
The disk morphology corresponding to these four epochs can be seen in Figure~\ref{fig:1} (panels {\it a}. {\it c}, {\it d} and {\it f}).
We confirmed that, at any epoch, the gas-pressure torque $N_{\rm th}$ does not significantly dominate either the magnetic torque $N_{\rm mag}$ or the gravitational torque $N_{\rm grav}$. It only rarely dominates other torques by a slight amount at certain radii, as shown for example by the blue dashed line in Figure~\ref{fig:11}.  
Since the contribution of the thermal torque is not great, we only focus on the magnetic and gravitational torques in the following discussion.

As the disk morphology evolves (see Fig.~\ref{fig:1}), the radial pattern of the ratio $N_{\rm mag}/N_{\rm grav}$ changes quantitatively, but not qualitatively, during the short duration after protostar formation.  
Figure~\ref{fig:11} indicates that although the ratio $N_{\rm mag}/N_{\rm grav}$ differs between epochs, the overall tendency remains the same. 
The ratio $N_{\rm mag}/N_{\rm grav}$ becomes low around $\sim1$\,AU, which roughly corresponds to the rotationally-supported disk where the magnetic field is very weak (the left panel of Fig.~\ref{fig:10}) and the spiral structure develops (Fig.~\ref{fig:2}).  
On the other hand, the magnetic torque dominates the gravitational torque in the inner ($\lesssim 0.3$\,AU) and outer ($\gtrsim3$\,AU) disk regions. 

In the magnetically inactive region, the angular momentum is not effectively transferred by the magnetic effects and the gas is supported by the centrifugal force against the gravity.
Therefore, in this region the surface density gradually increases and gravitational instability then occurs to form the spiral structure. The spiral structure is very effective at creating gravitational torques that then transfer the angular momentum, and a rapid mass accretion is induced. 
After the rapid mass accretion, the disk recovers a quiescent phase. 
The rapid accretion and quiescent phases are repeated and the episodic mass accretion is realized as seen in Figure~\ref{fig:9}. The episodic mass accretion also causes an episodic jet.
These episodes of higher accretion represent typically less than one order of magnitude rise in the accretion rate, and are similar to the low amplitude flickering seen in long-term simulations of gravitational instability driven evolution of protostellar disks \citep{vorobyov06,vorobyov10,vorobyov15}. However, the major bursts of mass accretion rate in those simulations that rise by several orders of magnitude are not seen in our simulation at these early evolutionary times.

\begin{figure}
\includegraphics[width=12cm]{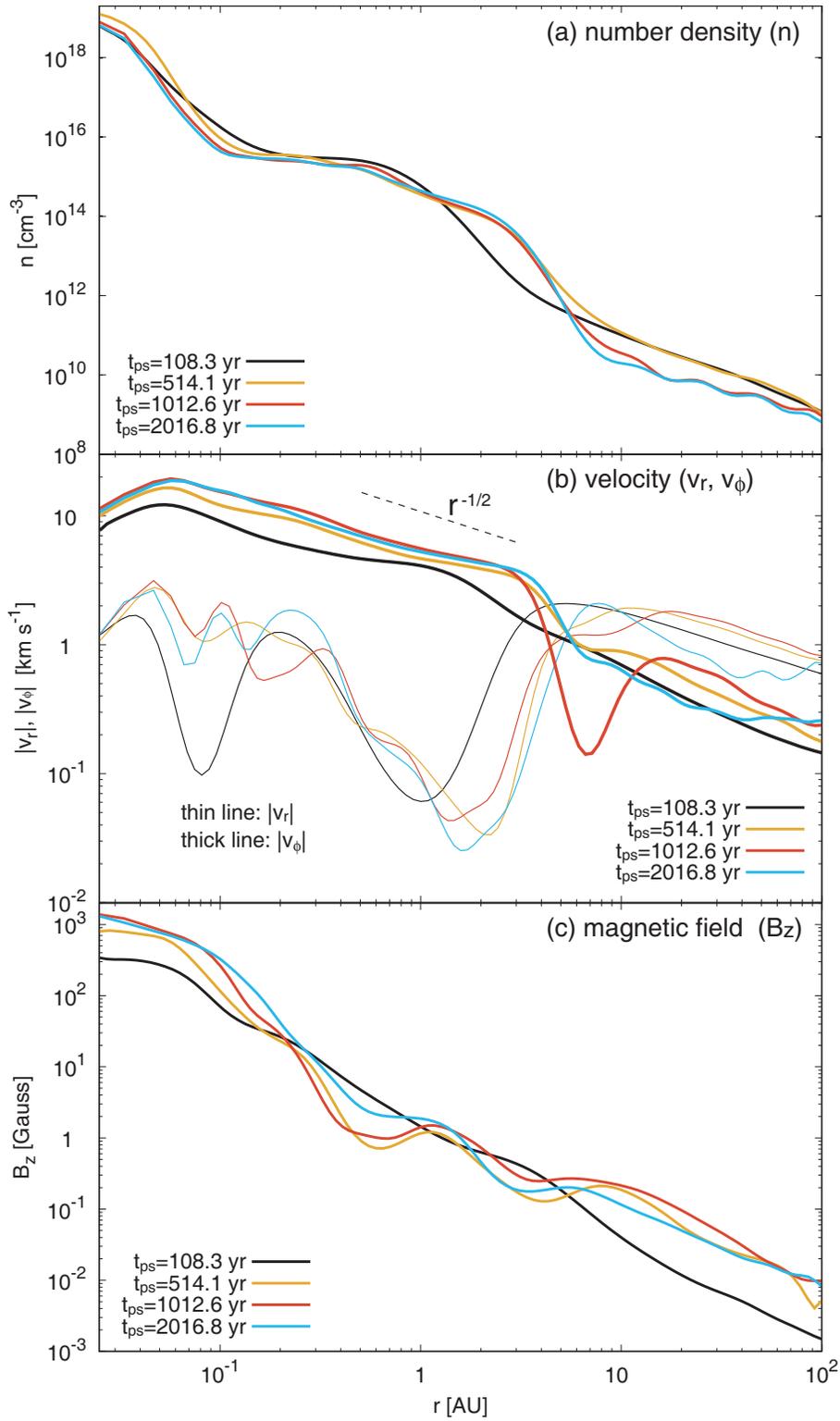}
\caption{
Density (top), velocity (middle), and $z$-component of magnetic field (bottom) on the equatorial plane at four different epochs are plotted against radial distance from the protostar.
}
\label{fig:12}
\end{figure}

Finally, we discuss the properties of the circumstellar region. 
In Figure~\ref{fig:12}, the azimuthally-averaged density, velocity and magnetic field profiles are plotted against the distance from the protostar. 
The density at each epoch suddenly increases around $r\sim5$\,AU, where also the radial velocity suddenly drops and the azimuthal velocity begins to increase (Fig.~\ref{fig:12}{\it a} and {\it b}). 
In addition, the azimuthal velocity is proportional to $\propto r^{-1/2}$ (Fig.~\ref{fig:12}{\it b}) in the range of 0.05\,AU $\lesssim r \lesssim$ 5\,AU.
Thus, we consider that a Keplerian disk exists in this region, which roughly corresponds to the disk-like structure seen in Figures~\ref{fig:1} and \ref{fig:2}. 
In addition, in Figure~\ref{fig:12}{\it c}, the magnetic field in the range of  $0.05\,{\rm AU}\lesssim r \lesssim 5\,{\rm AU}$ is not monotonically increasing as the radius decreases. 
This trend is caused by the Ohmic dissipation, which allows a rotationally-supported disk to form in this
magnetically inactive region where magnetic braking is shut off.

\subsection{Initial Condition and Cloud Parameters}
This study showed that the rotationally-supported disk forms and is sustained for 2000 years after protostar formation. 
In addition, both a low-velocity outflow and a high-velocity jet appear in the early phase of star formation. 
Our results qualitatively agree with high-resolution observations by ALMA \citep{sakai14,ohashi14,aso15,perez16,bjerkeli16,alves17,lee17,matsushita19}. 
However, since we only calculated the evolution of a single cloud core, we cannot conclude that the rotationally-supported disk, low-velocity outflow and high-velocity jet appear universally in the early star formation phase.

In this study, we have modeled a spherical cloud with a uniform magnetic field and rigid rotation as an initial state, in which typical values of the magnetic field strength and angular velocity were adopted as described in \S~\ref{sec:method}. 
However, prestellar cloud cores are expected to have a variety of magnetic field strengths \citep{crutcher99} and angular velocities \citep{caselli02}. 
Therefore, when we adopt a different cloud parameter (different magnetic field strength or different angular velocity), we may see a different evolutionary path of the star formation. 
For example, when the magnetic field is considerably strong and the initial cloud has a subcritical mass-to-flux ratio, the ambipolar diffusion will play a role in the early gas collapsing phase \citep{basu94,basu95a,basu95b} and the rotationally supported disk may not appear immediately after protostar formation \citep{machida18}.
In the future, we need to calculate the cloud evolution while varying the cloud parameters in order to fully clarify the star formation process.    
In addition, the initial state adopted in this study is highly idealized; the initial cloud has a spherical structure without turbulence and the uniform magnetic field vector is adopted to be parallel to the rotation axis. 
Here we have understood the star formation process in a simple setting as a first step.

\section{Summary}
\label{sec:summary}
Using a three-dimensional resistive MHD nested grid code, we calculated the cloud evolution from a prestellar cloud core until 2000 yr after protostar formation. We do not use a central sink and the protostar is resolved in our simulation. 
The following results are obtained:
\begin{itemize}
\item Formation of a rotationally supported disk following protostar formation.
The magnetic field dissipates in a high-density gas region where the magnetic braking is suppressed and a rotationally-supported disk forms. However, since the magnetic dissipation region is limited to a small area, the disk does not significantly grow and has a size of $\lesssim 4$\,AU in the very early phase.  

\item Episodic accretion due to disk gravitational instability.
Since the angular momentum is not efficiently transferred by the magnetic effects in the magnetic dissipation region, the infalling gas remains in the disk and increases the disk surface density. 
Then, the gravitational instability occurs to form the nonaxisymmetric structure that induces episodic accretion from the disk onto the protostar.

\item Time-variable high-velocity jet caused by the episodic mass accretion. 
The episodic mass accretion drives a time-variable high-velocity jet in the proximity of the protostar.  
The jet has a well collimated structure on the large scale, while the jet opening angle widens with time at its base. The jet velocity exceeds $100\km$,  which is roughly comparable to the Keplerian velocity at the surface of the protostar. 

\item Low-velocity outflow encloses the high-velocity jet.
In addition to the high-velocity jet, the low-velocity outflow is driven by the disk outer region.
Since the low-velocity outflow appears before the appearance of the high-velocity jet, the jet is embedded in the outflow. 
However, the jet catches up with the outflow in a short duration because the jet velocity is higher than the outflow velocity.
The outflow and jet create a cavity, shell, and knots that are frequently seen in observations.
The outflow mass, momentum, and kinetic energy of the low-velocity component dominates that of the high-velocity component. 
Thus, the low-velocity component (or low-velocity outflow) plays a significant role in the star formation process. 

\item Suppression of star formation efficiency by the outflow. 
A large fraction of cloud mass is blown away by the outflow (low-velocity outflow and high-velocity jet). 
At the end of the simulation, the outflow mass is three times greater than the protostellar mass, which means that the outflow may sufficiently suppress the star formation efficiency.
Although we need a further long-term simulations, the outflow may give an  impact on the star formation process. 
\end{itemize}

We compared our simulation results with observations in this study, both qualitatively and in some cases quantitatively.
Our simulation can explain many phenomena that are observed in the early phase of star formation.
However, a long-term time integration and/or parameter survey is necessary to comprehensively understand the role of magnetic fields in disk and outflow formation during star formation.

\section*{Acknowledgements}
We have benefited greatly from discussions with ~K. Tomida and ~T. Nakano. 
We also thank the referee for very useful comments on this paper. 
The present research used the computational resources of the HPCI system provided by (Cyber Sciencecenter, Tohoku University; Cybermedia Center, Osaka University, Earth Simulator, JAMSTEC) through the HPCI System Research Project (Project ID:hp160079, hp170047, hp180001,hp190035).
The present study was supported by JSPS KAKENHI Grant Numbers JP17K05387, JP17H02869,  JP17H06360 and  17KK0096.
Simulations reported in this paper were also performed by 2017 and 2018 Koubo Kadai on Earth Simulator (NEC SX-ACE) at JAMSTEC.
This work was partly achieved through the use of supercomputer system SX-ACE at the Cybermedia Center, Osaka University. SB was supported by a Discovery Grant from the Natural Sciences and Engineering Research Council of Canada.

\end{document}